\documentclass[nofootinbib,10pt,twocolumn,pra,showpacs,tightenlines, showkeys, floatfix, english,aps]{revtex4}
\usepackage[T1]{fontenc}
\usepackage[utf8]{inputenc}
\usepackage{color}
\usepackage{float}
\usepackage{epsfig,graphicx,times}
\usepackage{amstext}
\usepackage{amsmath}  
\usepackage{amssymb}
\usepackage{stackrel}
\usepackage{graphicx}
\usepackage{esint}
\usepackage{epstopdf}
\usepackage{epsfig}
\usepackage{bm}

\makeatletter

\DeclareFontEncoding{LGR}{}{}
\DeclareTextSymbol{\~}{LGR}{126}

\makeatother

\usepackage{babel}
\begin{document}

\title{{\Large{}Lower order and higher order entanglement in $87Rb$ $5S-5P-5D$
hyperfine manifold modeled as a four-wave mixing process}}

\author{{\normalsize{}Moumita Das$^{1}$, Biswajit Sen$^{2}$, }Ayan Ray$^{3}${\normalsize{}
and Anirban Pathak$^{4}$}}

\affiliation{$^{1}$Department of Physics, Siliguri College, Siliguri - 734 001,
India\\
$^{2}$Department of Physics, Vidyasagar Teachers' Training College,
Midnapore - 721 101, India\\
 $^{3}$Radiactive ion beam facilty group, Variable energy cyclotron
centre, 1/AF, Bidhan Nagar, Kolkata-700 064, India\\
$^{4}$ Jaypee Institute of Information Technology, A 10, Sector 62, Noida, UP 201307, India}

\begin{abstract}
{\normalsize{}Possibilities of generation of lower order and higher
order intermodal entanglement in }$87Rb$ $5S-5P-5D$ hyperfine manifold
{\normalsize{}are rigorously investigated using the Sen-Mandal perturbative
technique by showing the equivalence of the system with the four-wave
mixing (FWM) process. The investigation has revealed that for a set
of experimentally realizable parameters we can observe lower
order and higher order intermodal entanglement between pump and signal
modes and signal and idler modes in a FWM process
associated with the }$87Rb$ $5S-5P-5D$ hyperfine manifold{\normalsize{}.
In addition, trimodal entanglement involving pump, signal and idler
modes is also reported. }{\normalsize \par}
\end{abstract}
\maketitle

\section{Introduction\label{sec:Introduction}}

The phenomenon of entanglement has drawn considerable attention since
its inception in Einstein, Podolsky and Rosen's (EPR) thought experiment
\cite{einstein-epr}. Entanglement describes a system of particles
that have one or more highly correlated quantum properties, such as
position, momentum, spin, etc. Specifically, two sub-systems that
are entangled cannot be described by independent wavefunctions. Instead
the quantum state of the combined system is described by a single
wavefunction. For example, in the initial experiments on entanglement
conducted with optical parametric amplifier (OPA) \cite{hong-mandel1},
a linear intensity dependence was observed in the coincident absorption
probability. This has been explained \cite{hong-mandel1,Banacloche1,javanainen,fei}
by considering the absorption of the signal photon together with the
absorption of the idler counterpart, as they \textquoteleft \textquoteleft travel\textquoteright \textquoteright{}
together. This explanation is based on the correlated nature of two
photon state, though the usual quadratic dependence signature for
a two-photon absorption process is absent in the observation. However,
this is simple enough and deficient as it neither distinguishes between
the classical correlation and entanglement, nor does it say anything
about the higher order entanglement. Later, first part of the deficiency
was qualitatively addressed in Ref. \cite{muthukrishnan-agawral}.
Here, the time asymmetry, which is intrinsic to the two-photon state
vector produced by successive decay of a three-level cascade system,
is held responsible for the distinction between classical correlation
and entanglement. Still the order of entanglement obtainable from
a cascade system remains unanswered. In this paper, we have attempted
to address this problem by taking recourse to four wave mixing (FWM)
process in a cascade system. The FWM process intrinsically acts as
a generator of nonclassical states, especially as a generator of entangled
photons \cite{boyer} and offers a unique scope for analyzing the
existence of different orders of intermodal entanglement. In the quantum
description of the FWM process, simultaneous annihilation of two pump
photons (which may have different frequencies) creates a signal-idler
photon pair. 

Nonclassical properties associated with the FWM process have been
studied almost since the inception of quantum optics. In fact, squeezed
light was first experimentally generated using FWM \cite{sq-exp}.
In the last few decades, nonclassical properties associated with FWM
process have been studied in various ways (\cite{opo-using-FWM,freq-comb=000026SPD,superluminal-light-pulse}
and references therein). Applications of FWM have also been reported
in various contexts (\cite{opo-using-FWM,freq-comb=000026SPD,superluminal-light-pulse,spd1,spd2,optical filtering,low-noise-requency-conversion}
and references therein). Specifically, applications of FWM have been
reported for optical parametric oscillators (OPOs) \cite{opo-using-FWM},
frequency-comb sources \cite{freq-comb=000026SPD}, single photon
sources for quantum cryptography \cite{freq-comb=000026SPD,spd1,spd2},
stimulated generation of superluminal light pulses \cite{superluminal-light-pulse},
optical filtering \cite{optical filtering}, low noise chip-based
frequency converter \cite{low-noise-requency-conversion}, etc. Further,
in silicon nanophotonic waveguides, several useful optical phenomena
related to telecom-band ($\lambda\thickapprox1550\,nm)$ all-optical
functions (such as, wavelength conversion, signal regeneration and
tunable optical delay) have been demonstrated using FWM (see \cite{Liu}
and references therein). In addition to these, FWM has recently been
used to develop FWM microscopy \cite{FWM-microscopy}, which
is found to be very useful for the study of the nonlinear optical response
of nanostructures \cite{FWM-microscopy}; enhancement of FWM (i.e.,
larger value of third order susceptibility $\chi^{(3)}$ in comparison
to existing optical materials) has been observed in plasmonic nanocluster
\cite{Zheng}. 

Thus, we may comment that FWM is an extremely important process, which
acts as a test bed for studying non-classicality of photons 
\cite{opo-using-FWM,freq-comb=000026SPD,superluminal-light-pulse}.
This fact and the above mentioned applications have motivated us to
investigate a particular aspect of FWM for a cascade system: intermodal
entanglement. Specifically, in this paper, we investigate the possibilities
of generation of lower order and higher order entanglement in FWM
process associated with a cascade system because entanglement has
been established as one of the most important resource for quantum
information processing and quantum communication \cite{my book}.
To be precise, with the advent of quantum information theory, several
interesting phenomena (e.g., quantum teleportation \cite{Bennet1993},
dense coding \cite{densecoding}, etc.) are reported which do not
have any classical analogue and which require entanglement as an essential
resource. Consequently, several systems have already been investigated
as sources of entanglement (see \cite{pathak-perina,pathak-PRA} and
references therein). However, it is still interesting to find experimentally
realizable simple systems that can produce entanglement. In what follows,
we will show that FWM process associated with a cascade system can
provide us one such experimentally realizable and relatively simple
system. It would be apt to note that some efforts have already been
made to investigate the existence of intermodal entanglement in FWM
process, both theoretically and experimentally (\cite{ent-Glorieux,ent-Glorieux12,ent-Payne,ent-Wu,ent-Yu}
and references therein). However, to the best of our knowledge, higher
order entanglement is not studied in any of the existing works. Although,
studies on higher order nonclassicalities \cite{generalized-higher order,HOAwithMartin,Maria-PRA-1,Maria-2,higher-order-PRL,with Martin hammar}
have become relevant in the recent past. These works showed that there
is indeed a dire necessity to introduce a higher order nonclassical
criterion to detect weak nonclassicalities in a relatively easy
manner. Keeping these facts in mind, in the present paper, we investigate
the possibilities of observing lower order and higher order intermodal
entanglement in FWM process associated with a cascade system under
the framework of Sen-Mandal perturbative approach \cite{bsen1} that
is known to provide analytic expressions for time evolution of field
operators with greater accuracy compared to the traditionally used
short-time solution \cite{psgupta}. This is well established in earlier
works (\cite{pathak-PRA,kishore2014co-coupler,kishore2014contra,kishore2016-Zeno,pathak-pra2}
and references therein). In what follows, we report a perturbative
solution (using the Sen-Mandal approach) for the Heisenberg's equations
of motion for various modes present in the Hamiltonian of the FWM
process. The perturbative solution obtained here is subsequently
used to investigate the existence of lower order and higher order
entanglement using a set of inequalities that can be expressed as
moments of annihilation and creation operators. To be precise, we
have used here Duan et al.'s criterion \cite{duan} and Hillery Zubairy's
criteria \cite{HZ-PRL,HZ2007,HZ2010} to investigate the existence
of intermodal entanglement. The investigation has revealed the signatures
of the existence of lower order and higher order intermodal (two-mode)
entanglement for all possible combinations of modes (i.e., entanglement
is observed between (i) pump and idler modes, (ii) pump and signal
modes, (iii) idler and signal modes). Not only that the possibility
trimodal entanglement is also investigated here, and it is found that
the appropriate choice of parameters yields trimodal entanglement involving
pump, signal and idler modes. Remaining part of the paper is organized
as follows. In Section \ref{sec:The-cascade-level-system}, we describe
the relevance of using the $87Rb$ $5S-5P-5D$ hyperfine manifold
as a test bed for entanglement study by exercising FWM. The model
Hamiltonian for the FWM process described in Section \ref{sec:The-cascade-level-system},
is described in Section \ref{sec:The-model-Hamiltonian}. This takes
into consideration all the four modes as weak and thus quantum mechanical
and subsequently we report an operator solution of the Heisenberg's
equations of motion corresponding to the each mode of FWM process.
The solution is obtained using the Sen-Mandal perturbative approach. In
Section \ref{sec:Intermodal-entanglement}, possibilities of generation
of lower order and higher order (including tri-modal entanglement)
in FWM process are studied using the operator solutions obtained in
Section \ref{sec:The-model-Hamiltonian}. Finally, the paper is concluded
in Section \ref{sec:Conclusion}.

\section{The cascade level coupling scheme as test bed for entanglement\label{sec:The-cascade-level-system}}

A practical level scheme, where cascade decay can be observed, is
$87Rb$ $5S_{\frac{1}{2}}\leftarrow6P_{\frac{3}{2}}\leftarrow5D_{\frac{5}{2}}$
route (cf. Fig. \ref{fig:Scheme-of-FWM} a). The excitation of atoms
to $5D_{\frac{5}{2}}$ can be done in a two photon process either
through an intermediate state, i.e., (I) $5S_{\frac{1}{2}}\xrightarrow{780nm}5P_{\frac{3}{2}}\xrightarrow{776nm}5D_{\frac{5}{2}}$
\cite{kienlen} or through (II) a virtual level by using photons of
$778nm$ wavelength \cite{olson-meyer}. However (I) is a more practised
option to populate $5D_{\frac{5}{2}}$ as for (II) the absorption
cross-section is much smaller. Option (I) is used to describe several
important optical processes, e.g., FWM \cite{akulshin}, electromagnetically
induced transparency (EIT) \cite{banacloche-li}, double resonance
optical pumping (DROP) \cite{lee1}, optical switching \cite{ayan1},
etc. Fig. \ref{fig:Scheme-of-FWM} a illustrates the level scheme
under consideration where both (I) and (II) pathways are clearly shown.
The life times ($\tau$) are in order: $\tau_{5P_{\frac{3}{2}}}(26ns)<\tau_{6P_{\frac{3}{2}}}(112ns)<\tau_{DP_{\frac{5}{2}}}(240ns)$.
Hence $6P_{\frac{3}{2}}$ acts as a leaky reservoir $\left|r_{l}\right\rangle $
w.r.t., $5P_{\frac{5}{2}}$ with a leakage rate slow enough to satisfy
$\tau_{6P_{\frac{3}{2}}}\sim\tau_{5P_{\frac{3}{2}}}.$ So faster optical
pumping and decay cycles centered on $5P_{\frac{3}{2}}$ can be averaged
over a single $\tau_{6P_{\frac{3}{2}}}.$ This does not violate the
steady state condition; further $\left|r_{l}\right\rangle $ is decoupled
from the main system. So it bears negligible influence on cascade
excitations. On the other hand, $\tau_{6P_{\frac{3}{2}}}\sim0.5\tau_{5P_{\frac{3}{2}}}$;
hence spontaneous decay $5S_{\frac{1}{2}}\leftarrow6P_{\frac{3}{2}}\leftarrow5D_{\frac{5}{2}}$
would replicate hyperfine structure of $5D_{\frac{5}{2}}$ level (cf.
Fig. \ref{fig:Scheme-of-FWM} a). Similarly, Fig. \ref{fig:Scheme-of-FWM}
b illustrates a typical experimental spectrum obtained under pump-probe
Rabi frequency combination of $\Omega_{pu}(\Omega_{pr})\sim2\pi\times80MHz(2\pi\times2MHz).$
Both lasers are plane polarized and satisfy the $5S_{\frac{1}{2}}(F=2)\xrightarrow{{\rm Probe}}5P_{\frac{3}{2}}(F^{\prime}=3)\xrightarrow{{\rm Pump}}5D_{\frac{5}{2}}(F^{\prime\prime})$
connection. The X-axis of Fig. \ref{fig:Scheme-of-FWM} b (ii) is
calibrated with the saturation absorption spectrum of the probe laser
(inset; Fig. \ref{fig:Scheme-of-FWM} b (i)). Here the probe laser
is scanned and the pump is stationary. The spectrum shows prominent
signatures of DROP \cite{lee1,ayan1} on the Doppler broadened background
of probe absorption. This situation is further clarified when the probe
is locked to $F=2\rightarrow F^{\prime}=3$ and pump is scanned and
blue fluorescence of $5S_{\frac{1}{2}}(F=2)\xleftarrow{420nm}5P_{\frac{3}{2}}(F^{\prime}=3)\xleftarrow{5.23\mu m}5D_{\frac{5}{2}}(F^{\prime\prime})$
the decay channel is simultaneously monitored (see Fig. \ref{fig:Scheme-of-FWM}
b (iii), (iv)). In cases of Fig. \ref{fig:Scheme-of-FWM} b (ii),
(iv) the I, II, III represent DROP signals of $F=2\longrightarrow F^{\prime\prime}=4,\,3,\,2$
two photon transitions. Unlike Fig. \ref{fig:Scheme-of-FWM} b (ii),
the spectrum in (iv) has no Doppler background. Here the locked probe
laser in principle addresses small velocity groups of atoms resonant
with the same. These atoms further reach $\left|3\right\rangle $
by the pump laser itself. Due to participation of highly selective
velocity groups of atoms the resultant Doppler background is largely
reduced. Though the counter propagating pump-probe combination is
favorable for observing strong EIT under two photon resonance ($F=2\rightarrow F^{\prime}=3\rightarrow F^{\prime\prime}$)
condition (i.e., $\Delta_{pu}+\Delta_{pr}\approx0$ and $\Delta_{pu}\approx0\approx\Delta_{pr}$
); only trace of EIT is observed to be present at the tip of DROP
profiles. This may be explained by considering the decay route $F=2\xleftarrow{{\rm branching\,ratio\,}(\eta=1)}F^{\prime}=3\xleftarrow{\eta=0.76}F^{\prime\prime}$,\textcolor{red}{{}
}which forms a pseudo-closed absorption-emission cycle. However, due
to velocity selective nature (considering $\sqrt{\Omega^{2}+\Delta^{2}}$
is the generalized Rabi frequency where, $\Delta$ is laser detuning)
of optical pumping, it is apparent that there are also other decay
routes, which remain active, resulting in a sufficient population of
$F=1$ state facilitating DROP condition \cite{lee2} and the EIT
is almost obscured . The striking feature of Fig. 1B (iii), i.e. the
blue fluorescence, is its one-to-one correspondence with the two photon
absorption. This is because blue photons mainly originate from $5S_{\frac{1}{2}}\xleftarrow{\eta=0.23}6P_{\frac{3}{2}}\xleftarrow{\eta=0.26}F^{\prime\prime}$
\cite{noh} decay channel. In a simplistic manner we may think that
under relatively higher pump power $\left(I_{pump}\geq I_{saturation\,5P_{\frac{3}{2}}\rightarrow5D_{\frac{5}{2}}}\right)$\textcolor{red}{{}
}and relatively lower probe power $\left(I_{probe}\leq I_{saturation\,5S_{\frac{1}{2}}\rightarrow5P_{\frac{3}{2}}}\right)$,\textcolor{red}{{}
}the blue light intensity ($I_{Blue}$) bears a correlation of $I_{Blue}\infty I_{probe}.$
It shows that there exists finite possibility to produce spontaneous
emission in a cascade decay, which at least remain intensity correlated
to the probe. This intrinsic capability of cascade emission deserves
further attention to explore if any kind of phase correlation is obtainable.
To generate phase correlation, the co-propagating pump-probe configuration
holds edge over the counter-propagating one. This is because FWM is
realizable under the first kind of alignment where the phase matching
condition demands: $\bar{k}_{780nm}+\bar{k}_{776nm}=\bar{k}_{5.2\mu m}+\bar{k}_{420nm}.$
This indeed offers an unique scope where the phase and intensity correlation
can be simultaneously obtained. By definition, two beams of light
can be made quantum mechanically entangled through correlations of
their phase and intensity fluctuations. Hence the cascade system under
co-propagating laser action may act as a source of non-classical light. 

For a two photon cascade decay$\left(\left|3\right\rangle \rightarrow\left|2\right\rangle \rightarrow\left|1\right\rangle \right)$
the situation may be understood as follows: (i) initially the atom
occupies highest excited state $\left|3\right\rangle $ while the
field remains in vacuum $\left|0\right\rangle ,$ (ii) the excited
atom decays to intermediate state $\left|2\right\rangle $ by emitting
a photon and makes a final dash to $\left|1\right\rangle $ with emission
of another photon. To consider the total process, it is not possible
to distinguish between the photon emission sequences like $\left|3\right\rangle \rightarrow\left|2\right\rangle $
followed by $\left|2\right\rangle \rightarrow\left|1\right\rangle $
or vice versa. In such situation, indistinguishablity of photon comes
into play. For a cascade system, the product state for (i) is $\left|3;0\right\rangle =\left|3\right\rangle \left|0\right\rangle $
while for (ii) they are $\left|2;1_{\bar{k}s}\right\rangle =\left|2\right\rangle \left|1_{\bar{k}s}\right\rangle $
and $\left|1;1_{\bar{k}s,}1_{\bar{k^{\prime}s}}\right\rangle =\left|1\right\rangle \left|1_{\bar{k}s,}1_{\bar{k^{\prime}s}}\right\rangle .$
The total state vector becomes a linear superposition of these individual
state vectors. At the very beginning the atom and the field remains
entangled. But at later times the decoherence of the excited level
population destroys the entanglement. As a result the final product
state appears as: $\left|3;1_{\bar{k}s,}1_{\bar{k^{\prime}s}}\right\rangle =\left|3\right\rangle \left|1_{\bar{k}s,}1_{\bar{k^{\prime}s}}\right\rangle $.
In this case $\left|1_{\bar{k}s,}1_{\bar{k^{\prime}}s\prime}\right\rangle ,$
which is a two photon state, may be considered to be an entangled
state if it is not separable (i.e., $\bar{k}s\neq\bar{k^{\prime}}s^{\prime}$
); where are wave vector (polarization) of radiation. For a cascade
medium the degree of entanglement is determined by the ratio of linewidths
of upper and intermediate excited states \cite{muthukrishnan-agawral,scully-book}.
 Since in the cascade emission of Fig. \ref{fig:Scheme-of-FWM} a
$\tau_{5D}>\tau_{6P}$; it is in principle possible to generate entangled
light through non-degenerate FWM process.

\section{{\normalsize{}The model Hamiltonian\label{sec:The-model-Hamiltonian}}}
\begin{widetext}

\begin{figure}
\centering{}\emph{\includegraphics[scale=0.85]{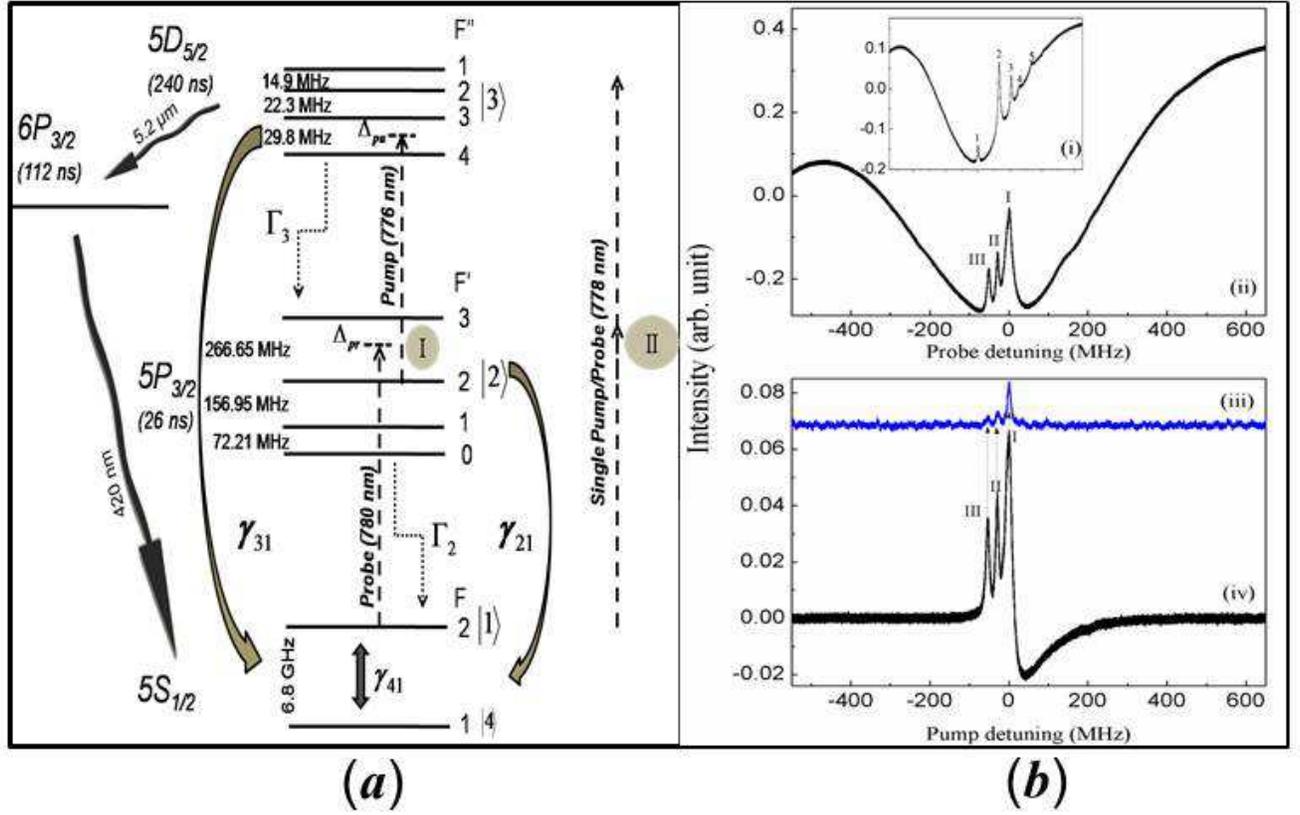}}\protect\caption{\label{fig:Scheme-of-FWM}(Color online) Level scheme with result for counter propagating
wave. \textbf{(a)} Level scheme in $5S_{\frac{1}{2}}\rightarrow5P_{\frac{3}{2}}(D_{2})\rightarrow5D_{\frac{5}{2}}$
transition (I) of Rubidium atom ($^{87}Rb$), relevant for DROP (EIT)
experiments. The pump (probe) laser beams are linearly polarized.
Dephasing between ground state ($\left|1\right\rangle $ ) hyperfine
components ($F=2,\,1$) is $\gamma_{g,\,1\leftrightarrow4}$, governed
by the transit time broadening. Spontaneous decay rate from $\left|j\right\rangle $
is $\Gamma_{j}$ (dotted arrows). Here, $\Gamma_{3}=2\pi\times0.97\,MHz,$
$\Gamma_{2}=2\pi\times6.066\,MHz,$ $\Gamma_{1}=0$ are the natural
linewidths. The coherent dephasing rate $\left|j\right\rangle \rightarrow\left|i\right\rangle $
is $\gamma_{ji}\thickapprox\left(\Gamma_{j}+\Gamma_{i}\right)/2.$
The decay route $5S_{\frac{1}{2}}\rightarrow6P_{\frac{3}{2}}\left\{{\rm{lifetime}}\sim112ns\right\} \leftarrow5D_{\frac{5}{2}}$
is used to monitor population history at $5D_{\frac{5}{2}}.$ Lifetimes
of other states are also mentioned below respective level captions.
The direct excitation $5S_{\frac{1}{2}}\rightarrow5D_{\frac{5}{2}}$
(II) is also possible through virtual level. \textbf{(b)} Here, the
saturation absorption spectrum (i) of Rb is shown where 1,4,5 correspond
to $F=2\rightarrow F^{\prime}=3,\,2,\,1$ hyperfine components, whereas
2,3 correspond to crossover transitions $F=2\rightarrow F^{\prime}=3,\,2$
and $3,1$. The DROP spectra $F=2\rightarrow F^{\prime\prime}=4,\,3,\,2$
are presented under (ii) probe laser frequency scan (pump stationary)
with Doppler background and (iv) pump laser detuning (probe static)
without Doppler background. The plot in (iii) presents the blue fluorescence
spectra mimicking the hyperfine separation of $5D_{\frac{5}{2}}$
state.}
\end{figure}

\end{widetext}

The quantum mechanical Hamiltonian for FWM process shown in Fig. \ref{fig:Scheme-of-FWM}
is 

\begin{equation}
\begin{array}{lcl}
H & = & \omega_{a}a^{\dagger}a+\omega_{b}b^{\dagger}b+\omega_{c}c^{\dagger}c+g\left(a^{2}b^{\dagger}c^{\dagger}+a^{\dagger2}bc\right),\end{array}\label{hamil.}
\end{equation}
where $g$ is the interaction constant and the $a(a^{\dagger}),$
$b(b^{\dagger})$ and $c(c^{\dagger})$ are annihilation (creation)
operators for two degenerate pump modes, signal mode and idler mode,
respectively. Now from the Fig. \ref{fig:Scheme-of-FWM} $a$, $\left|1\right\rangle $
- $\left|3\right\rangle $ is populated using two-photon process with
the pump mode $a$ and the transition from $\left|3\right\rangle $
to $6P_{3/2}$ represents the mode $b$ of the Hamiltonian (\ref{hamil.}).
Finally, mode $c$ corresponds to the transition from $6P_{3/2}$ to $5S_{1/2}.$
Here we consider all the modes as weak and that requires a completely
quantum mechanical treatment. To obtain the time evolution of the
annihilation operators of different modes, we first obtain the Heisenberg's
equations of motion for various field operators as

\begin{equation}
\begin{array}{lcl}
\overset{.}{a}\left(t\right) & = & -i\left(\omega_{a}a+2ga^{\dagger}bc\right)\\
\overset{.}{b}\left(t\right) & = & -i\left(\omega_{b}b+ga^{2}c^{\dagger}\right)\\
\overset{.}{c}\left(t\right) & = & -i\left(\omega_{c}c+ga^{2}b^{\dagger}\right)
\end{array}.\label{EOM}
\end{equation}
These equations are coupled, nonlinear differential equations of field
operators and are not exactly solvable in closed analytical forms.
Consequently, it is required that we follow a perturbative approach.
Here, we have used the Sen-Mandal perturbative technique \cite{pathak-PRA,bsen1,pathak-pra2},
which is already known to be more general than the well-known short-time
approximation approach \cite{psgupta}. Now, following
Sen-Mandal's perturbative technique, we can write assumed solutions
(assumed analytic forms of the the time evolution of annihilation
operators of various modes) as

\begin{equation}
\begin{array}{lcl}
a\left(t\right) & = & f_{1}a+f_{2}a^{\dagger}bc+f_{3}ab^{\dagger}bc^{\dagger}c+f_{4}a^{\dagger}a^{2}c^{\dagger}c+f_{5}a^{\dagger}a^{2}bb^{\dagger},\\
b\left(t\right) & = & g_{1}b+g_{2}a^{2}c^{\dagger}+g_{3}a^{2}a^{\dagger2}b+g_{4}a^{\dagger}abcc^{\dagger}+g_{5}aa^{\dagger}bcc^{\dagger},\\
c\left(t\right) & = & h_{1}c+h_{2}a^{2}b^{\dagger}+h_{3}a^{2}a^{\dagger2}c+h_{4}a^{\dagger}acbb^{\dagger}+h_{5}aa^{\dagger}cbb^{\dagger},
\end{array}\label{assumedsolution}
\end{equation}
where $f_{i},$$g_{i}$ and $h_{i}s$ are time dependent parameters.

The above mentioned assumed solution is obtained by using the fact
that the time evolution of the annihilation operator $a(t)$ under
Hamiltonian $H$ can be expressed as 
\begin{equation}
\begin{array}{lcl}
a\left(t\right) & = & \exp\left(iHt\right)a\left(0\right)\exp\left(-iHt\right),\end{array}\label{timeevolution}
\end{equation}
and the same (i.e., Eq. (\ref{timeevolution})), can be expanded as
\begin{equation}
\begin{array}{lcl}
a\left(t\right) & = & a\left(0\right)+it\left[H,a\left(0\right)\right]+\frac{\left(it\right)^{2}}{2!}\left[H,\left[H,a\left(0\right)\right]\right]\\
&+&\frac{\left(it\right)^{3}}{3!}\left[H,\left[H,\left[H,a\left(0\right)\right]\right]\right]+\cdots.\end{array}\label{eq:taylorseries}
\end{equation}
It is to be noted that we have neglected the terms beyond $g^{2}$, but have not imposed any restriction on time $t$ provided
$gt\ll1$. Specifically, the assumed solution is
obtained by keeping all the terms that arise from the infinite series
(\ref{eq:taylorseries}), provided that the terms are not of higher
power (higher than quadratic) in $g$.  Subsequently, the
assumed solution for a specific mode is substituted in the Heisenberg's
equation of motion for that particular mode which is obtained using
the given Hamiltonian. Now, the coefficients of the similar terms
are compared to obtain a set of coupled ordinary differential equations
involving $f_{i},\,g_{i},$ and  $h_{i}$. Finally, this set of coupled
differential equations is solved to obtain the final analytic solution.
This process leads to some additional terms that are not obtained
in the conventional short-time solution.\textcolor{red}{{} }These extra
terms provide an edge to the Sen-Mandal method in comparison to the
conventional short-time method. Now, following the prescription described
above in general and using Eqns. (\ref{EOM}) and (\ref{assumedsolution})
in particular, we can obtain the functional forms of these parameters
as

\begin{equation}
\begin{array}{lcl}
f_{1} & = & e^{-i\omega_{a}t}\\
f_{2} & = & \frac{2g}{\triangle\omega_{1}}f_{1}\left(1-e^{i\triangle\omega_{1}t}\right)\\
f_{3} & = & \frac{2g}{\triangle\omega_{1}}\left(f_{2}+i2gtf_{1}\right)\\
f_{5} & = & f_{4}=-\frac{f_{3}}{2}
\end{array}\label{coeff. of a}
\end{equation}

\begin{equation}
\begin{array}{lcl}
g_{1} & = & e^{-i\omega_{b}t}\\
g_{2} & = & -\frac{g}{\triangle\omega_{1}}g_{1}\left(1-e^{-i\triangle\omega_{1}t}\right)\\
g_{3} & = & -\frac{g}{\triangle\omega_{1}}\left(g_{2}+igtg_{1}\right)\\
g_{5} & = & g_{4}=-2g_{3}
\end{array}\label{coeff. of b}
\end{equation}

\begin{equation}
\begin{array}{lcl}
h_{1} & = & e^{-i\omega_{c}t}\\
h_{2} & = & -\frac{g}{\triangle\omega_{1}}h_{1}\left(1-e^{-i\triangle\omega_{1}t}\right)\\
h_{3} & = & -\frac{g}{\triangle\omega_{1}}\left(h_{2}+igth_{1}\right)\\
h_{5} & = & h_{4}=-2h_{3},
\end{array}\label{coeff. of c}
\end{equation}
where\textcolor{green}{{} }$\Delta\omega_{1}=2\omega_{a}-\omega_{b}-\omega_{c}$.
Thus, we obtain a perturbative solution for the equations of motion
corresponding to the Hamiltonian of the FWM process. The correctness
of the above solutions can can be checked by the equal time commutation
relation (ETCR) i.e., by verifying that $[a(t),\,a^{\dagger}(t)]=[b(t),\,b^{\dagger}(t)]=[b(t),\,b^{\dagger}(t)]=1.$
The obtained solutions may now be used to investigate the existence
of lower order and higher order entanglement involving various modes
by using a set of moment-based criteria of entanglement. The same
is done in the following section.

\section{{\normalsize{}Intermodal entanglement\label{sec:Intermodal-entanglement}}}

We have already mentioned that entanglement plays very crucial role
in quantum information processing. There exist several inseparability
criteria \cite{duan,HZ-PRL,HZ2007,HZ2010,Adam-criterion} which may
be used to investigate the possibility of the existence of entanglement
in the physical systems. Many of these inseparability criteria (e.g.,
Hillery and Zubairy's criteria \cite{HZ-PRL,HZ2007,HZ2010}, Duan
et al.'s criterion \cite{duan}, etc.) are moment based (i.e., they
are expressed in terms moments of the annihilation and creation operators
of two or more modes), and thus suitable for the present study as
we already have closed form analytic expressions for the time evolution
of various modes. Interestingly, most of these inseparability criteria
are only sufficient and not necessary. Thus, if inseparability condition
is found to satisfy, we know with certainty that the investigated
state is entangled, but if the condition is not satisfied we cannot
conclude anything about the separability. Keeping this in mind, we
usually investigate the existence of entanglement using two or more
inseparability criteria, so that if one criterion fails to detect
entanglement for a specific state and specific parameters, the other
criterion (criteria) may succeed to detect it. In what follows, we
will use two criteria of Hillery and Zubairy and one criterion of
Duan et al. Further, we would like to note that these inseparability
criteria can be classified as: (i) Lower order criteria: If an inseparability
criterion (inequality) contains terms only up to fourth order in the annihilation
and/or creation operators of different modes, the criterion is referred
to as a lower order criterion. This is so, as to study the correlations
between two modes we need at least fourth order terms. Consequently,
all criteria that involve terms up to 4th order in annihilation/creation
operators are known as lower order criteria. (ii) Higher order criteria:
If an inseparability criterion (inequality) contain terms of order
higher than the fourth order in the annihilation and/or creation operators
of different modes, the criterion is referred to as a higher order criterion.
We can easily see that by this definition all inseparability criteria
for three or more modes (or involving three or more particles) must
be higher order criteria. Interestingly, one can also construct higher
order criteria for the investigation of entanglement in two-mode case
\cite{HZ-PRL,Adam-criterion}. In what follows, we will study higher
order entanglement from both the perspectives (i.e., three-mode cases
and two mode higher order cases), but to begin with, in the next subsection
we investigate the possibility of observing lower order entanglement
between two modes of FWM process.

To investigate the possibilities of observing entanglement in FWM
process, we consider that the initial state $|\psi(0)\rangle$ is
separable and is the product of three coherent states corresponding to
three modes of the system. Further, we assume that $|\alpha|^{2},\,|\beta|^{2}$
and $|\gamma|^{2}$ are the initial number of photons in each pump
mode, signal mode and idler mode, respectively. Thus, we have 
\begin{equation}
\begin{array}{lcl}
|\psi(0)\rangle & = & \left|\alpha\beta\gamma\right\rangle =\left|\alpha\right\rangle \otimes\left|\beta\right\rangle \otimes\left|\gamma\right\rangle \end{array}.\label{compositecoherentstate}
\end{equation}
\textcolor{blue}{{} }

\subsection{{\normalsize{}Lower order two-mode entanglement}}

To obtain the signature of entanglement in FWM process, we have used
the following inseparability criteria introduced by Hillery and Zubairy
\cite{HZ-PRL,HZ2007,HZ2010} 
\begin{equation}
\begin{array}{lcl}
E_{a,b} & = & \left\langle N_{a}N_{b}\right\rangle -\left|\left\langle ab^{\dagger}\right\rangle \right|^{2}<0\end{array}\label{hz1critreria}
\end{equation}
and 
\begin{equation}
\begin{array}{lcl}
E_{a,b}^{\prime} & = & \left\langle N_{a}\right\rangle \left\langle N_{b}\right\rangle -\left|\left\langle ab\right\rangle \right|^{2}<0\end{array}\label{hz2criteria}
\end{equation}
where $a$ and $b$ represent two arbitrary modes. Throughout our
present paper, we refer to these inequalities (\ref{hz1critreria})
and (\ref{hz2criteria}) as HZ1 and HZ2 criteria, respectively. As
these two criteria are only sufficient not necessary, we also use another
moment based inseparability criterion which is referred to as Duan
et al.'s criterion \cite{duan} for any two arbitrary mode $a$ and
$b$, Duan et al.'s criterion describes the condition of inseparability
as 
\begin{equation}
D_{ab}=\begin{array}{lcl}
\left(\Delta u\right)^{2}+\left(\Delta v\right)^{2}-2 & < & 0\end{array},\label{duanetalcriteria}
\end{equation}
 where
\begin{equation}
\begin{array}{lcl}
u & = & \frac{1}{\sqrt{2}}\left\{ \left(a+a^{\dagger}\right)+\left(b+b^{\dagger}\right)\right\} ,\\
v & = & -\frac{i}{\sqrt{2}}\left\{ \left(a-a^{\dagger}\right)+\left(b-b^{\dagger}\right)\right\} .
\end{array}
\end{equation}
Now using Eqns. (\ref{assumedsolution}), (\ref{compositecoherentstate})
and HZ1 criterion (\ref{hz1critreria}), we obtain

\begin{equation}
\begin{array}{lcl}
E_{a,b} & = & \left|f_{2}\right|^{2}\left(\frac{1}{4}\left|\alpha\right|^{6}+\left|\beta\right|^{4}\left|\gamma\right|^{2}-\frac{1}{2}\left|\alpha\right|^{4}\left|\beta\right|^{2}-\left|\alpha\right|^{2}\left|\beta\right|^{2}\left|\gamma\right|^{2}\right),\end{array}\label{ab_hz1}
\end{equation}

\begin{equation}
\begin{array}{lcl}
E_{b,c} & = & \left|g_{2}\right|^{2}\left[\left|\alpha\right|^{4}\left(1+3\left|\beta\right|^{2}+3\left|\gamma\right|^{2}\right)-2\left|\beta\right|^{2}\left|\gamma\right|^{2}\right.\\
&\times&\left.\left(1+2\left|\alpha\right|^{2}\right)\right]+\left(h_{1}h_{2}^{*}\alpha^{*2}\beta\gamma+{\rm c.c.}\right)
\end{array},\label{bc-hz1}
\end{equation}

\begin{equation}
\begin{array}{lcl}
E_{a,c} & = & \left|f_{2}\right|^{2}\left(\frac{1}{4}\left|\alpha\right|^{6}+\left|\beta\right|^{2}\left|\gamma\right|^{4}-\frac{1}{2}\left|\alpha\right|^{4}\left|\gamma\right|^{2}-\left|\alpha\right|^{2}\left|\beta\right|^{2}\left|\gamma\right|^{2}\right)\end{array},\label{ac_hz1}
\end{equation}

\begin{widetext}

\begin{figure}
\includegraphics[angle=-0,scale=0.5]{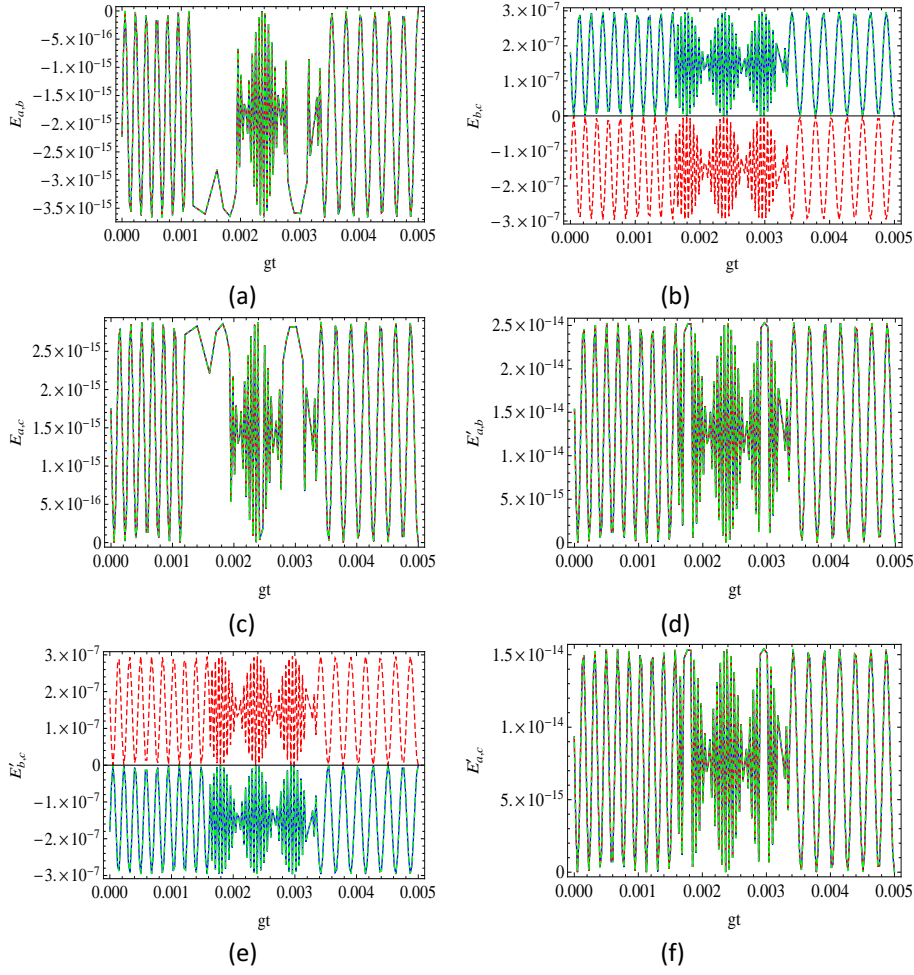}

\protect\caption{\label{fig:Lower-order-HZ1 and HZ2}(Color online) Lower order entanglement
using HZ1 and HZ2 criteria. The solid (blue), dash-dotted (green),
and dashed (red) lines represent the phase angle of the input complex
amplitude $\alpha_{s}$ for $\phi=0$, $\pi/2$ and $\pi$, respectively,
using $\omega_{a}=242.38\times10^{13}$ Hz,  $\omega_{b}=36.05\times10^{13}$ Hz, 
$\omega_{c}=448.98\times10^{13}$ Hz, $\alpha=5,$ $\beta=4$ and $\gamma=2.$
Intermodal entanglement is observed using HZ1 criterion in the coupled
mode: (a) $ab$ mode for all values of the phase angle $\phi,$ (b) $bc$ mode for
$\phi=\frac{\pi}{2}$. No signature of entanglement in (c) $ac$ mode
is observed using HZ1 criterion, but using HZ2 criterion, we observed entanglement in (e)
$bc$ mode for the phase angle $0$ and $\pi$ only. No signature
of inter modal entanglement is observed in (d) $ab$ mode and (e)
ac mode using HZ2 criterion. }
\end{figure}

\end{widetext}

\flushleft where we have assumed that the initial state is described by (\ref{compositecoherentstate}).
The same initial state is used in the entire paper. Now, we plot right
hand sides of (\ref{ab_hz1})-(\ref{ac_hz1}) in Fig. \ref{fig:Lower-order-HZ1 and HZ2}(a)-(c)
with three values of the phase $\phi$ of input pump mode $a$. Precisely,
we have considered $\alpha=|\alpha|\exp(i\phi)$ and plotted right
hand sides of (\ref{ab_hz1})-(\ref{ac_hz1}) using $\phi=0,\frac{\pi}{2}$
and $\pi$. Negative regions of the plots clearly illustrate bi-modal
entanglement in $ab$ modes for all values $\phi$ and in $bc$ modes
for $\phi=\frac{\pi}{2}$. We could not find any signature of entanglement
for $ac$ mode. However, we cannot conclude anything about the separability/inseparability
in those cases where negative regions are not found. This is so because
the HZ1 criterion and other similar criteria of inseparability used
in this paper are only sufficient and not necessary. Similarly, we
may use (\ref{assumedsolution}), (\ref{compositecoherentstate})
and (\ref{hz2criteria}) to obtain

\begin{widetext}

\begin{equation}
\begin{array}{lcl}
E_{a,b}^{\prime} & = & \left|f_{2}\right|^{2}\left(\frac{1}{4}\left|\alpha\right|^{6}+\left|\beta\right|^{4}\left|\gamma\right|^{2}+\frac{1}{2}\left|\alpha\right|^{4}\left|\beta\right|^{2}+\left|\alpha\right|^{2}\left|\beta\right|^{2}\left|\gamma\right|^{2}\right)\end{array},\label{ab_hz2}
\end{equation}

\begin{equation}
\begin{array}{lcl}
E_{b,c}^{\prime} & = & \left|g_{2}\right|^{2}\left[2\left|\beta\right|^{2}\left|\gamma\right|^{2}\left(1+2\left|\alpha\right|^{2}\right)-\left|\alpha\right|^{4}\left(1+\left|\beta\right|^{2}+\left|\gamma\right|^{2}\right)\right]- \left(h_{1}h_{2}^{*}\alpha^{*2}\beta\gamma+c.c.\right)
\end{array},\label{bc_hz2}
\end{equation}

\begin{equation}
\begin{array}{lcl}
E_{a,c}^{\prime} & = & \left|f_{2}\right|^{2}\left(\frac{1}{4}\left|\alpha\right|^{6}+\left|\beta\right|^{2}\left|\gamma\right|^{4}+\frac{1}{2}\left|\alpha\right|^{4}\left|\gamma\right|^{2}+\left|\alpha\right|^{2}\left|\beta\right|^{2}\left|\gamma\right|^{2}\right).\end{array}\label{ac_hz2}
\end{equation}

\end{widetext}

As before, we plot right hand sides of Eqns. (\ref{ab_hz2})-(\ref{ac_hz2})
in Figs.\ref{fig:Lower-order-HZ1 and HZ2}(d)-(f) and it is clear
from the figures that the intermodal entanglement is observed only
in $bc$ mode for the phase angle $\phi=0\,{\rm and}\,\pi.$ No signature
of intermodal entanglement is observed in the remaining two cases.
As discussed above, in these cases we are nonconclusive about the
inseparability.

Now, we may extend our investigation on inseparability using Duan
et al.'s criterion described in Eq. (\ref{duanetalcriteria}) and
obtain the following expressions with the help of the solutions (\ref{assumedsolution}),
(\ref{coeff. of a})-(\ref{coeff. of c}) described in the previous
section:

\begin{equation}
\begin{array}{lcl}
D_{ab} & = & D_{ac}=\bigl|f_{2}\bigr|^{2}\left(\frac{1}{2}\bigl|\alpha\bigl|^{4}+2\bigl|\beta\bigl|^{2}\bigr|\gamma\bigr|^{2}\right),\end{array}\label{ab_duan}
\end{equation}

\begin{equation}
\begin{array}{lcl}
D_{bc} & = & \bigl|f_{2}\bigr|^{2}\bigl|\alpha\bigr|^{4}\end{array},\label{bc_duan}
\end{equation}
Right hand sides of Eqns. (\ref{ab_duan}) - (\ref{bc_duan}) are
clearly positive and thus we may conclude that for the physical system
studied here, Duan et. al.'s criterion cannot identify any signature
of entanglement. \textcolor{red}{{} }

\subsection{{\normalsize{}Higher order entanglement}}

In order to investigate the existence of bi-modal higher order entanglement
in FWM process, we may use the following two criteria introduced by
Hillery and Zubairy \cite{HZ-PRL}: 
\begin{equation}
\begin{array}{lcl}
E_{a,b}^{m,n} & = & \left\langle a^{\dagger m}a^{m}b^{\dagger n}b^{n}\right\rangle -\left|\left\langle a^{m}b^{\dagger n}\right\rangle \right|^{2}<0\\
E_{a,b}^{\prime m,n} & = & \left\langle a^{\dagger m}a^{m}\Bigr\rangle\Bigl\langle b^{\dagger n}b^{n}\right\rangle -\left|\left\langle a^{m}b^{n}\right\rangle \right|^{2}<0,
\end{array}
\end{equation}
where $a$ and $b$ are two arbitrary modes and $m$ and $n$ are
the positive integers. Here, $m+n\geq3$ gives the criteria for higher
order entanglement. The negativity of the right hand side would show
the signature of the entanglement. Now, using the first higher order
criteria of Hillery and Zubairy for various field modes we obtain

\begin{widetext}

\begin{equation}
\begin{array}{lcl}
E_{a,b}^{m,n} & = & \left|f_{2}\right|^{2}\left[m^{2}\left|\alpha\right|^{2m-2}\left|\beta\right|^{2n+2}\left|\gamma\right|^{2}+m^{2}n\left|\alpha\right|^{2m-2}\left|\beta\right|^{2n}\left|\gamma\right|^{2}\right.\\
 & - & \frac{m^{2}n(m+1)}{2}\left|\alpha\right|^{2m-2}\left|\beta\right|^{2n}\left|\gamma\right|^{2}-mn\left|\alpha\right|^{2m}\left|\beta\right|^{2n}\left|\gamma\right|^{2}\\
 & - & \frac{mn}{2}\left|\alpha\right|^{2m+2}\left|\beta\right|^{2n}-\frac{m^{2}n(m-1)^{2}}{4}\left|\alpha\right|^{2m-4}\left|\beta\right|^{2n}\left|\gamma\right|^{2}\\
 & + & \left.\begin{array}{c}
\frac{n^{2}}{4}\left|\alpha\right|^{2m+4}\left|\beta\right|^{2n-2}\end{array}-\frac{mn(m-1)}{4}\left|\alpha\right|^{2m}\left|\beta\right|^{2n}\right],
\end{array}\label{HZ1abmn}
\end{equation}
\begin{equation}
{\color{red}\begin{array}{lcl}
{\normalcolor E_{b,c}^{m,n}} & {\normalcolor =} & {\normalcolor \left|g_{2}\right|^{2}\left\{ \left(2mn^{2}+n^{2}\right)\left|\alpha\right|^{4}\left|\beta\right|^{2m}\left|\gamma\right|^{2n-2}+m^{2}n^{2}\left|\alpha\right|^{4}\left|\beta\right|^{2m-2}\left|\gamma\right|^{2n-2}\right.}\\
{\normalcolor } & {\normalcolor +} & {\normalcolor \left.\left(2m^{2}n+m^{2}\right)\left|\alpha\right|^{4}\left|\beta\right|^{2m-2}\left|\gamma\right|^{2n}-2mn\left(1+2\left|\alpha\right|^{2}\right)\left|\beta\right|^{2m}\left|\gamma\right|^{2n}\right\} }\\
{\normalcolor } & {\normalcolor +} & {\normalcolor \left[\left\{ mn\Bigl(h_{1}h_{2}^{*}\frac{\alpha^{*2}}{\beta^{*}\gamma^{*}}+\left(m-1\right)\frac{h_{2}^{2}}{h_{1}^{2}}\frac{\alpha^{4}\beta^{*}}{\beta\gamma^{2}}\Bigr)\left|\beta\right|^{2m}\left|\gamma\right|^{2n}+mn\left(m-1\right)\frac{h_{2}^{2}}{h_{1}^{2}}\right.\right.}\\
{\normalcolor } & {\normalcolor \times} & {\normalcolor {\normalcolor \left.\left.\Bigl(\frac{\alpha^{4}\gamma^{*}}{\beta^{2}\gamma}+\frac{\left(n-1\right)}{2}\frac{\alpha^{4}}{\beta^{2}\gamma^{2}}\Bigr)\left|\beta\right|^{2m}\left|\gamma\right|^{2n}\right\} +{\rm c.c.}\right]},}
\end{array}}\label{eq:hz1bcmn}
\end{equation}
and

\begin{equation}
\begin{array}{lcl}
E_{a,c}^{m,n} & = & \left|f_{2}\right|^{2}\left[m^{2}\left|\alpha\right|^{2m-2}\left|\gamma\right|^{2n+2}\left|\beta\right|^{2}+m^{2}n\left|\alpha\right|^{2m-2}\left|\gamma\right|^{2n}\left|\beta\right|^{2}\right.\\
 & - & \frac{m^{2}n(m+1)}{2}\left|\alpha\right|^{2m-2}\left|\gamma\right|^{2n}\left|\beta\right|^{2}-mn\left|\alpha\right|^{2m}\left|\gamma\right|^{2n}\left|\beta\right|^{2}\\
 & - & \frac{mn}{2}\left|\alpha\right|^{2m+2}\left|\gamma\right|^{2n}-\frac{m^{2}n(m-1)^{2}}{4}\left|\alpha\right|^{2m-4}\left|\gamma\right|^{2n}\left|\beta\right|^{2}\\
 & + & \left.\begin{array}{c}
\frac{n^{2}}{4}\left|\alpha\right|^{2m+4}\left|\gamma\right|^{2n-2}\end{array}-\frac{mn(m-1)}{4}\left|\alpha\right|^{2m}\left|\gamma\right|^{2n}\right].
\end{array}\label{eq:HZ1acmn}
\end{equation}

\end{widetext}

\begin{figure}
\includegraphics[angle=0,scale=0.38]{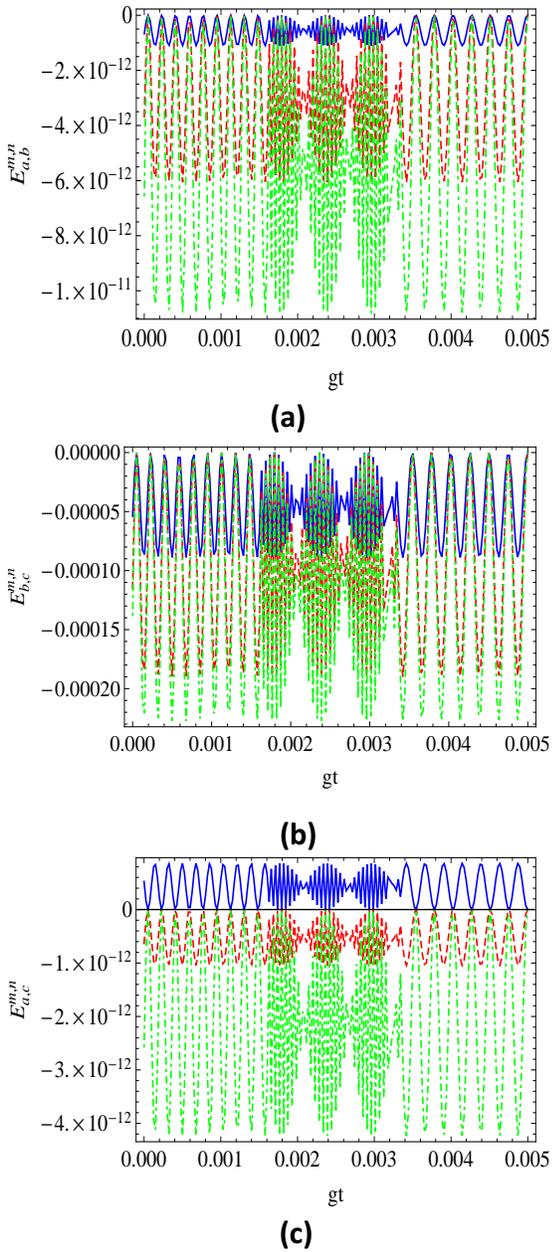}

\protect\caption{\label{fig:Hz1higher order} (Color online) Plot of higher order entanglement using
HZ1 criterion using $\omega_{a}=242.38\times10^{13}$ Hz, $\omega_{b}=36.05\times10^{13}$ Hz,
$\omega_{c}=448.98\times10^{13}$ Hz, $\alpha=5,$ $\beta=4$ and $\gamma=2.$ The solid (blue) line, dashed (red) line and dash-dotted (green)
line represent the $n=1$, and $m=1,$ $2$ and $3$, respectively. Here,
in all the plots $n=1$ and $m=2$ and $3$ are multiplied by 300 and 20,
respectively. Higher order intermodal entanglement is observed in (a)
ab mode, (b) bc mode for phase angle $\phi=\frac{\pi}{2}$ and for
(c) ac mode. }
\end{figure}

\begin{figure}
\includegraphics[angle=0,scale=0.38]{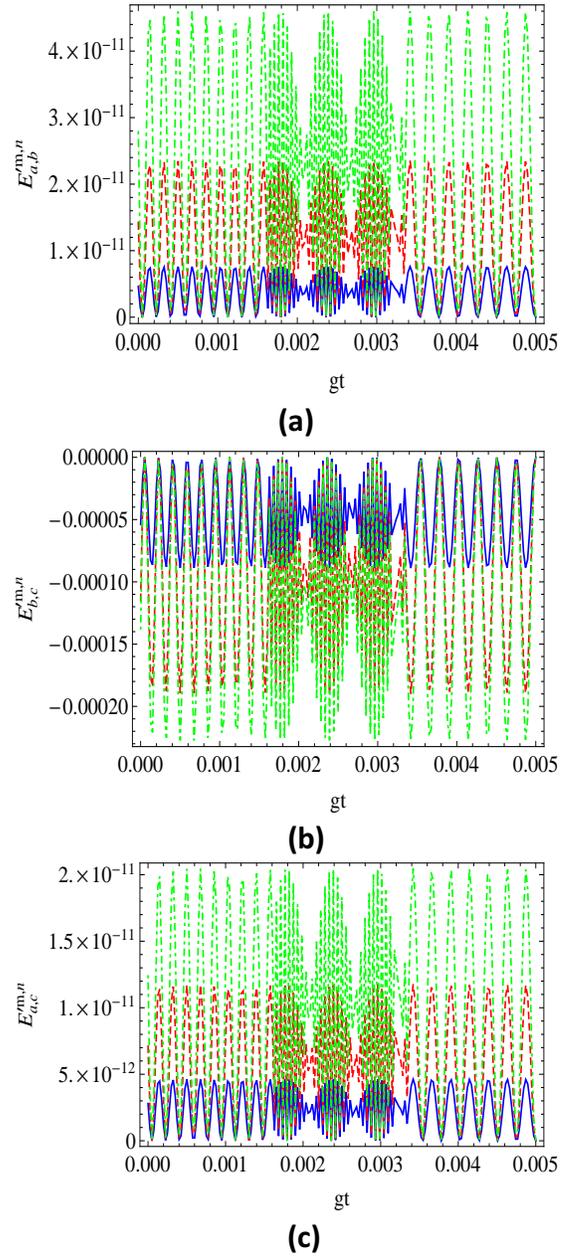}

\protect\caption{\label{fig:hz2higherorder} (Color online) Higher order entanglement using HZ2 criterion
using $\omega_{a}=242.38\times10^{13}$ Hz, $\omega_{b}=36.05\times10^{13}$ Hz, 
$\omega_{c}=448.98\times10^{13}$ Hz, $\alpha=5,$ $\beta=4$ and $\gamma=2.$ The solid (blue), dashed (red), and dash-dotted (green) lines
represent $n=1$ and $m=1,$ $2$ and $3$, respectively. Here,
in figure, $m=1$ and $n=2$ and $3$ are multiplied by 300 and 20,
respectively. Higher order intermodal entanglement is observed in
(b) $bc$ mode for phase angle $0$ and $\pi$ (not shown) only; but
no signature of intermodal entanglement is observed in (a) $ab$ and 
(c) $ac$ modes.}
\end{figure}

To illustrate the fact that Eqs. (\ref{HZ1abmn})-(\ref{eq:HZ1acmn})
provide us signatures of higher order entanglement, we have plotted
the right hand sides of these equations with the dimensionless interaction
time $gt,$ in Fig. \ref{fig:Hz1higher order} a-c, where negative
regions of the curves depict the existence of higher order entanglement.
These figure reveal that the signature of the higher order entanglement
is observed in all three possible combinations in which two modes
can be chosen. Here, in Fig.\ref{fig:Hz1higher order}, there are
three lines in each plot, and they represent three cases for each
choice of modes i.e., for $n=1$ and $m=1,$$2$ and $3$ respectively.
Plot for $m=n=1$ (i.e., blue smooth line) corresponds to normal order
entanglement, whereas the $n=1,\,m=2$ (red dashed line) and $n=1,\,m=3$
(green dot dashed line) correspond to the higher order entanglement.
It is clear from the figure that the higher order intermodal entanglement
is observed for all three modes using HZ1 criterion. Further, Fig.\ref{fig:Hz1higher order}
c illustrates that for $ac$ modes signature of lower order entanglement
is not observed, but that of higher order entanglement is observed.
In the similar manner, using the second criteria of the Hillery-Zubeiry
we obtain

\begin{widetext}

\begin{equation}
\begin{array}{lcl}
E_{a,b}^{\prime m,n} & = & \left|f_{2}\right|^{2}\left[m^{2}\left|\alpha\right|^{2m-2}\left|\beta\right|^{2n+2}\left|\gamma\right|^{2}+\frac{1}{4}mn\left(m-1\right)\left|\alpha\right|^{2m}\left|\beta\right|^{2n}\right.\\
 & + & \left.\frac{mn}{2}\left|\alpha\right|^{2m+2}\left|\beta\right|^{2n}+\begin{array}{c}
\frac{n^{2}}{4}\left|\alpha\right|^{2m+4}\left|\beta\right|^{2n-2}\end{array}+mn\left|\alpha\right|^{2m}\left|\beta\right|^{2n}\left|\gamma\right|^{2}\right],
\end{array}
\end{equation}
\begin{equation}
{\color{red}\begin{array}{lcl}
{\normalcolor E_{b,c}^{\prime m,n}} & {\normalcolor =} & {\normalcolor \left|g_{2}\right|\left\{ \left(m^{2}-2m^{2}n\right)\left|\alpha\right|^{4}\left|\beta\right|^{2m-2}\left|\gamma\right|^{2n}-m^{2}n^{2}\left|\alpha\right|^{4}\left|\beta\right|^{2m-2}\left|\gamma\right|^{2n-2}\right.}\\
{\normalcolor } & {\normalcolor +} & {\normalcolor \left.\left(1-2m\right)n^{2}\left|\alpha\right|^{4}\left|\beta\right|^{2m}\left|\gamma\right|^{2n-2}+2mn\left(1+2\left|\alpha\right|^{2}\right)\left|\beta\right|^{2m}\left|\gamma\right|^{2n}\right\} }\\
{\normalcolor } & {\normalcolor -} & {\normalcolor \left[\left\{ mn\Bigl(h_{1}h_{2}^{*}\frac{\alpha^{*2}}{\beta^{*}\gamma^{*}}+\left(m-1\right)\frac{h_{2}^{2}}{h_{1}^{2}}\frac{\alpha^{4}\gamma*}{\beta^{2}\gamma}\Bigr)\left|\beta\right|^{2m}\left|\gamma\right|^{2n}\right.\right.}\\
{\normalcolor } & {\normalcolor +} & {\normalcolor \left.\left.mn\left(n-1\right)\frac{h_{2}^{2}}{h_{1}^{2}}\Bigl(\frac{\alpha^{4}\beta^{*}}{\beta\gamma^{2}}+\frac{(m-1)}{2}\frac{\alpha^{4}}{\beta^{2}\gamma^{2}}\Bigr)\left|\beta\right|^{2m}\left|\gamma\right|^{2n}\right\} +{\rm c.c.}\right]}{\normalcolor ,}
\end{array}}
\end{equation}
and
\begin{equation}
\begin{array}{lcl}
E_{a,c}^{\prime m,n} & = & \left|f_{2}\right|^{2}\left[m^{2}\left|\alpha\right|^{2m-2}\left|\beta\right|^{2}\left|\gamma\right|^{2n+2}+\begin{array}{c}
\frac{n^{2}}{4}\left|\alpha\right|^{2m+4}\left|\gamma\right|^{2n-2}\end{array}+\frac{mn}{2}\right.\\
 & \times & \left.\left|\alpha\right|^{2m+2}\left|\gamma\right|^{2n}+\frac{mm(m-1)}{4}\left|\alpha\right|^{2m}\left|\gamma\right|^{2n}+mn\left|\alpha\right|^{2m}\left|\beta\right|^{2}\left|\gamma\right|^{2n}\right]
\end{array}.
\end{equation}

\end{widetext}

Right hand sides of the above set of equations are plotted in Fig. \ref{fig:hz2higherorder}
a-c, which clearly show the signature of the higher order intermodal
entanglement for $bc$ mode for the phase angle $0$ and $\pi$. However, it does not show the same for other choices of modes. Thus,
HZ2 criterion could not detect the signature of higher order entanglement
in $ab$ and $ac$ mode. In other words, HZ2 criterion fails to detect
the higher order intermodal entanglement for $ab$ and $ac$ modes
for any phase angle $\phi$. 

There is another way to investigate the higher order entanglement.
Any multi-mode entangled state (which involve more than two modes)
is considered to be higher order entangled. In what follows, we investigate
the possibility of observing trimodal entanglement in FWM process
using the following criterion \cite{Ent condition-multimode} 
\begin{equation}
\begin{array}{lcl}
E_{a,b,c} & = & \left\langle N_{a}N_{b}N_{c}\right\rangle -\left|\left\langle abc^{\dagger}\right\rangle \right|^{2}<0\end{array},
\end{equation}
where $a,\,b,\,c$ represents three different modes. Now, we may note
that above criterion actually contain three criteria which represent
three different bipartite cuts, and we may obtain analytic expressions
for $E_{i,j,k}$ for various choices of $k$ mode as follows:

\begin{widetext}

\begin{equation}
\begin{array}{lcl}
E_{a,b,c} & = & \left|f_{2}\right|^{2}\left\{ \frac{1}{4}\left|\alpha\right|^{6}\left(1+3\left|\beta\right|^{2}+3\left|\gamma\right|^{2}\right)-\frac{1}{2}\left|\alpha\right|^{2}\left|\beta\right|^{2}\left|\gamma\right|^{2}\right.\\
 & \times & \left.\left(5\left|\alpha\right|^{2}+2\left|\beta\right|^{2}+3\right)+\left|\beta\right|^{4}\left|\gamma\right|^{4}\right\} \\
 & + & \left\{ \left(h_{1}h_{2}^{*}\left|\alpha\right|^{2}\alpha^{*2}\beta\gamma+f_{1}^{*}f_{2}g_{1}g_{2}^{*}\alpha^{*4}\beta^{2}\gamma^{2}\right)+{\rm c.c.}\right\} ,
\end{array}\label{abc_hz1}
\end{equation}

\begin{flushleft}
\begin{equation}
\begin{array}{lcl}
E_{b,c,a} & = & \left|f_{2}\right|^{2}\left\{ \frac{1}{4}\left|\alpha\right|^{6}\left(\left|\beta\right|^{2}+\left|\gamma\right|^{2}\right)-\left(1+\left|\alpha\right|^{2}+\left|\beta\right|^{2}+\left|\gamma\right|^{2}\right)\right.\\
 & \times & \left.\left|\alpha\right|^{2}\left|\beta\right|^{2}\left|\gamma\right|^{2}+\left|\beta\right|^{4}\left|\gamma\right|^{4}\right\} ,
\end{array}\label{bca_hz1}
\end{equation}
and
\par\end{flushleft}

\begin{equation}
\begin{array}{lcl}
E_{a,c,b} & = & \left|f_{2}\right|^{2}\left\{ \frac{1}{4}\left|\alpha\right|^{6}\left(1+3\left|\beta\right|^{2}+3\left|\gamma\right|^{2}\right)-\frac{1}{2}\bigl|\alpha\bigr|^{2}\left|\beta\right|^{2}\left|\gamma\right|^{2}\right.\\
 & \times & \left.\left(5\left|\alpha\right|^{2}+2\bigl|\gamma\bigr|^{2}+3\right)+\bigl|\beta\bigr|^{4}\left|\gamma\right|^{4}\right\} \\
 & + & \left\{ \left(h_{1}h_{2}^{*}\left|\alpha\right|^{2}\alpha^{*2}\beta\gamma+f_{1}^{*}f_{2}g_{1}g_{2}^{*}\alpha^{*4}\beta^{2}\gamma^{2}\right)+{\rm c.c.}\right\} .
\end{array}\label{acb_hz1}
\end{equation}

\end{widetext}

\begin{widetext}

\begin{figure}
\includegraphics[angle=-90,scale=0.7]{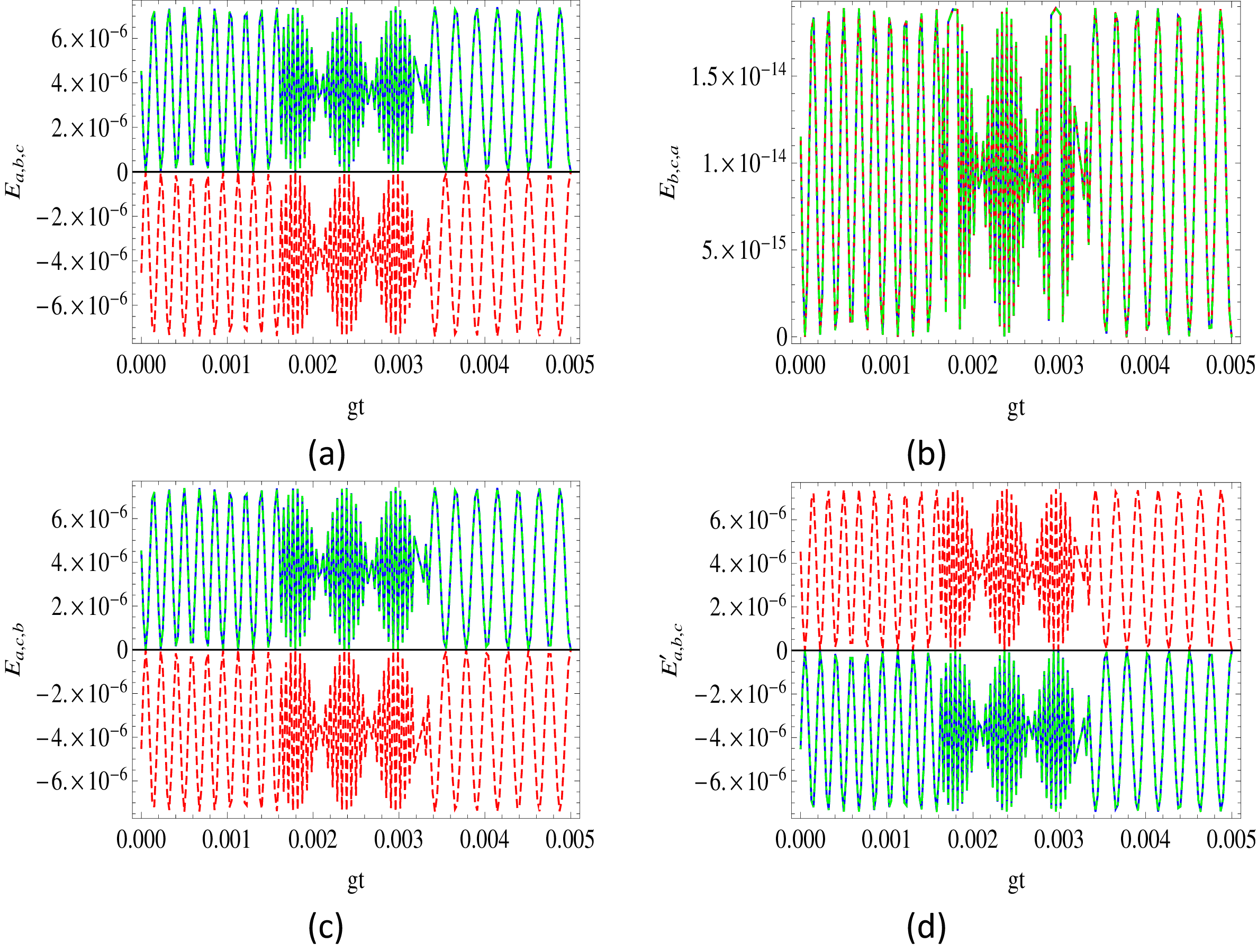}

\protect\caption{\label{fig:Tri-partaite-entanglement} (Color online)  Tri-partaite entanglement using
the HZ1 and HZ2 criteria.The solid (blue), dash-dotted (green)
and dashed (red) lines represent the phase angle of the input complex
amplitude $\alpha_{s}$ for $\phi=0$, $\pi/2$ and $\pi$, respectively,
using $\omega_{a}=242.38\times10^{13}$ Hz, $\omega_{b}=36.05\times10^{13}$ Hz,
$\omega_{c}=448.98\times10^{13}$ Hz, $\alpha=5,$ $\beta=4$ and $\gamma=2.$
Trimodal entanglement is observed using HZ1 criterion in (a) $abc$ mode for
phase angle $\phi=\frac{\pi}{2}$ (c) $acb$ mode and for the phase angle
$\phi=\frac{\pi}{2}.$ No signature of entanglement is observed in (c) $bca$
mode. (d) Trimodal entanglement using HZ2 is observed for the phase
angle $\phi=0$ and $\pi$ only.}
\end{figure}

\end{widetext}

If any of the above quantity (i.e., $E_{abc},\,E_{acb},\,E_{bca}$)
is found to be negative, we will have a signature of trimodal entanglement.
To obtain this signature Eqs. (\ref{abc_hz1})-(\ref{acb_hz1}) are
plotted in Fig. \ref{fig:Tri-partaite-entanglement} a-c, where we
observe negative regions for appropriate choices of phase. This indicates
the existence of the trimodal entanglement. To further illustrate
the existence of trimodal entanglement involving pump, signal and
idler modes, we further investigate its existence by using a symmetric
criterion for inseparability of three modes. The criterion describes
the condition for trimodal entanglement as $\left\langle N_{a}\right\rangle \left\langle N_{b}\right\rangle \left\langle N_{c}\right\rangle -\left|\left\langle abc\right\rangle \right|^{2}<0$
and using Eqs. (\ref{assumedsolution}) and (\ref{compositecoherentstate}),
we obtian

\begin{widetext}

\begin{equation}
\begin{array}{lcl}
\left\langle N_{a}\right\rangle \left\langle N_{b}\right\rangle \left\langle N_{c}\right\rangle -\left|\left\langle abc\right\rangle \right|^{2} & = & \left|f_{2}\right|^{2}\left\{ -\frac{1}{4}\left|\alpha\right|^{6}\left(1+\left|\beta\right|^{2}+\left|\gamma\right|^{2}\right)+\left|\alpha\right|^{2}\left|\beta\right|^{2}\left|\gamma\right|^{2}\right.\\
 & \times & \left.\left(3\left|\alpha\right|^{2}+\left|\beta\right|^{2}+\left|\gamma\right|^{2}+\frac{3}{2}\right)+\left|\beta\right|^{4}\left|\gamma\right|^{4}\right\} \\
 & - & \left\{ \left(h_{1}h_{2}^{*}\left|\alpha\right|^{2}\alpha^{*2}\beta\gamma+f_{1}f_{2}^{*}h_{1}^{*}h_{2}\alpha^{4}\beta^{*2}\gamma^{*2}\right)+{\rm c.c.}\right\} .
\end{array}\label{abc_hz2}
\end{equation}

\end{widetext}

Variation of RHS of this particular equation is plotted in Fig. \ref{fig:Tri-partaite-entanglement}
d, which also illustrate the possible existence of trimodal entanglement
(for specific choice phase) via its negative region.

\section{{\normalsize{}Conclusion\label{sec:Conclusion}}}

Traditionally, FWM process is viewed as a third order nonlinear optical
phenomenon having applications in various fields as summarized in
Sec. \ref{sec:Introduction}. Recently, the domain of the generation and
applicability of FWM have been considerably amplified. Specifically,
several new applications of FWM have been proposed. On the other hand,
applications of entangled states have been reported in various areas
of quantum information. Motivated by these facts, we have rigorously
investigated the generation of lower order and higher order intermodal
entanglement in FWM process using a set of moment-based criteria (criteria
based on moments of annihilation and creation operators) and a physical
system illustrated in Fig.\ref{fig:Scheme-of-FWM}, where FWM happens.
To be precise, present discussion is focused on a cascade system shown
in Fig.\ref{fig:Scheme-of-FWM}, where FWM process occurs for $87Rb$
$5S-5P-5D$ hyperfine manifold. Considering experimentally achievable
parameters, it's observed that the parameters whose negative values
indicate the existence of higher/lower order entanglement show an
oscillatory nature with variation of rescaled time. Further, it is
observed that suitable choice of the phase of the initial coherent
states plays a crucial role in the generation of entanglement. Here,
we observed bimodal entanglement between pump and signal modes and
signal and idler modes, and have also observed trimodal entanglement
using all these modes. Thus, the output of FWM process studied here
appears to be highly entangled. We conclude the paper with a hope
that this highly entangled output of FWM process would find some applications
in quantum communication and/or quantum information processing as
FWM is a process that can be easily realized experimentally (cf. Fig.\ref{fig:Scheme-of-FWM})
and as entanglement is one of the most important resource for quantum
communication and computation.

\textbf{Acknowledgment: }AP thanks Department of Science and Technology
(DST), India for the support provided through the project number EMR/2015/000393. .
The authors also thank Kishore Thapliyal for his interest in the work
and some fruitful discussions.

\end{document}